# Fabrication process-induced variations of Nb/Al/$AlO_x$/Nb Josephson junctions in superconductor integrated circuits

**Sergey K Tolpygo[1,2,3] and Denis Amparo[2]**

[1] HYPRES, Inc., 175 Clearbrook Rd., Elmsford, NY 10523, USA
[2] Department of Physics and Astronomy, Stony Brook University, Stony Brook, NY 11794, USA
[3] Department of Electrical and Computer Engineering, Stony Brook University, NY 11794, USA

E-mail: stolpygo@hypres.com; denis.amparo@stonybrook.edu

**Abstract.** Currently, superconductor digital integrated circuits fabricated at HYPRES, Inc. can operate at clock frequencies of ~ 40 GHz. The circuits present multilayered structures containing tens of thousands of Nb/Al/$AlO_x$/Nb Josephson junctions (JJs) of various sizes interconnected by four Nb wiring layers, resistors, and other circuit elements. In order to be operational, the integrated circuits should be fabricated such that the critical currents of Josephson junctions remain within the tight design margins and the proper relationships between the critical currents of JJs of different sizes are preserved. We present experimental data and discuss mechanisms of process-induced variations of the critical current and energy gap of Nb/Al/$AlO_x$/Nb Josephson junctions in integrated circuits. We demonstrate that the Josephson critical current may depend on the type and area of circuit elements connected to the junction, on the circuit pattern, and on the step in the fabrication process at which the connection is made. In particular, we discuss the influence of **a)** junction base electrode connection to ground plane, **b)** junction counter electrode connection to ground plane, and **c)** counter electrode connection to Ti/Au or Ti/Pd/Au contact pads by Nb wiring. We show that the process-induced changes of the properties of Nb/Al/$AlO_x$/Nb junctions are caused by migration of impurity atoms (hydrogen) between different layers comprising the integrated circuits.

## 1. Introduction

Superconductor integrated circuits based on Rapid Single Flux Quantum (RSFQ) logic can operate at very high clock frequencies and with very low energy dissipation. For instance, Nb-based integrated circuits that we currently fabricate at HYPRES, Inc. have clock frequencies of about 40 GHz. They utilize Nb/Al/$AlO_x$/Nb Josephson junctions (JJs) with $j_c$ = 4.5 kA/cm$^2$ critical current density [1], [2]. A large number of complex digital circuits containing over $10^4$ JJs have recently been fabricated with this technology for several cryocooler-based superconductor electronic systems designed and made at HYPRES [3].

It is well known that the maximum clock frequency of RSFQ circuits scales nearly proportionally to $j_c^{1/2}$, or to $1/a$, where $a$ is the typical junction size (see, e.g., [1]). Therefore, further increasing the clock frequency requires higher-$j_c$ junctions of smaller sizes. Fabrication processes with $j_c$ = 10 kA/cm$^2$ ($a$=1 μm) and 20 kA/cm$^2$ ($a$=0.8 μm) are currently under development in ISTEC, Japan and HYPRES,

Inc., respectively [1],[4]. The complexity and functionality of RSFQ circuits increase with the number of logic cells, depending mainly on the linewidth of inductors, resistors, and transmission lines connecting the cells, and only weakly on the size of JJs, because they currently occupy only a small fraction of circuit area. In addition to decreasing the linewidth, the integration scale and functionality can be increased by increasing the number of available physical layers in the fabrication process. In [1] we pointed out the existence of a trade-off between the speed and complexity of RSFQ circuits, namely that the maximum clock frequency generally decreases as the number of junctions in circuits increases. It remains to be seen if this trend continues to hold also for future generations of fabrication technologies.

At present, functionality of superconductor digital circuits is limited by a relatively small size of yieldable circuits. This size can be characterized by the typical number of junctions in fabricated circuits which are operational at high clock frequencies with reasonable probability. Currently this number is $\sim 2\cdot 10^4$. This yield limitation is a result of sensitivity of RSFQ circuits to deviations of the critical current of the junctions, $I_c$, from the design values, which reduce the margin of circuit operation and the maximum clock frequency. If the critical current deviations were caused by small random fluctuations of junction sizes in the fabrication process, the size of yieldable circuits could be much larger even with the present fabrication technologies (up to $\sim 10^5$ JJs); it would be mainly determined by statistical fluctuations of the critical current of the smallest Josephson junctions in the circuits [5]. However, it has been our experience that relatively large systematic deviations of the critical currents of Josephson junctions in integrated circuits often take place and put severe restrictions on the yield of complex circuits. Because elimination of these deviations is very important for the progress of superconductor digital electronics, we have recently undertaken a systematic study of all possible sources of process-induced variations of the critical current of Josephson junctions in superconductor integrated circuits [7]-[9]. The present paper continues this study. After a brief overview, we present new results on the fabrication process-induced variability of Josephson junctions in integrated circuits. For the obvious practical reasons, we concentrate on Nb-based circuits with Nb/Al/$AlO_x$/Nb junctions, and HYPRES fabrication process [1],[2],[6].

## 2. Systematic deviations of the critical current of Josephson junctions with ground connection

In superconductor integrated circuits, Josephson junctions are interconnected by superconducting wiring layers; all bias currents and signals are returned to the common ground plane. In the HYPRES fabrication process [1], there are five superconducting layers: ground plane (layer M0), first wiring layer M1 [it is also a part of junction base electrode (BE)], junction counter electrode (CE), and two wiring layers M2 and M3. Connections to the junction electrodes are made through contact holes in interlayer insulators I0 (between M0 and M1), I1B (between CE or M1 and M2), and I2 (between M2 and M3). Depending on the logic cell design, some junctions have a low inductance superconducting connection between BE and M0 using M1 wiring (see, e.g., Fig. 1 in [8]), some have a connection between CE and M0 via M2 wiring (see, e.g. Fig. in []), and some are not directly connected to the ground plane but are separated from it by other junctions and/or resistors. With respect to this ground plane connection we call these junctions as M1-GND, M2-GND, and floating, respectively.

In [7] we found several types of systematic differences between $I_c$s of nominally identical junctions in integrated circuits. Firstly, the critical current of grounded JJs is higher than the critical current of floating JJs. The difference is much larger than the typical statistic spread of critical currents in junctions of the same size. Secondly, there is a systematic difference between junctions with ground connection to the base electrode (M1-GND) and to the counter electrode (M2-GND). Both effects vary in size from run to run and across process wafers, indicating that they are process related.

We proposed a possible mechanism for higher $j_c$ in M1-GND JJs than in floating ones based on the fact that electric currents can flow through M1-GND junctions during plasma processing steps of fabrication when the counter electrode is exposed to processing plasma. These plasma-induced currents can cause electric stress (damage) to the tunnel barrier, whereas no such currents can flow through the floating junctions [7],[8]. However, this explanation does not work for M2-GND junctions because no significant current can flow through the tunnel barrier in these junctions (except in a few very special cases, which realization in a systematic manner is highly unlikely). In [8] we also found that the increase in $I_c$ of M1-GND junctions depends on the shape and area of the ground plane. We found a decrease in damage in M1-GND junctions with increasing the resistance of the ground plane connection in a qualitative agreement with the plasma damage model. However, quantitative agreement required assuming unusually high plasma potential nonuniformities which are very unlikely to exist in modern plasma processing systems if not created intentionally.

In [10] we directly measured the effect of dc electric stress on the critical current and properties of tunnel barriers in Nb/Al/$AlO_x$/Nb junctions with initial barrier transparencies similar to those used in integrated circuits. We found that, due to a high transparency, an irreversible degradation of the tunnel barrier and increase in $I_c$ start at stress current densities of a few mA/$\mu m^2$. Such currents cannot be supplied by our plasma processing tools that typically provide plasma current densities of only a few mA/$cm^2$. Therefore, plasma current damage cannot cause the increase of $I_c$ in many junctions on the wafer due to insufficient current. In the worst case, plasma damage perhaps can happen only in a small number of junctions connected to very large "antennae" – conductive structures that can collect charges from a very large area and channel them through the junctions [11]. The ratio of "antenna" area to the junction area should be on the order of $10^8$ for this mechanism to be operational for the presently used tunnel barriers and plasma processing conditions.

In [9] we proposed that the cause of the increased critical current in M1-GND junctions with respect to floating junctions is chemical or electrochemical in nature and is consistent with a diffusion and/or electromigration of impurities between the base electrode and the ground plane through contact holes in I0 insulator. The most likely diffusing substance is hydrogen which can dissolve in Nb in large quantities and has very high mobility at room temperatures [12]. The effect of hydrogen inclusion on the critical current in Nb/$AlO_x$/Nb junctions was observed in [13].

In our previous works [7]-[10] we were mainly concerned with variations of the critical current of Josephson junctions in integrated circuits because they are detrimental for the circuits' performance. However, a change in the critical current reflects changes in both the average transparency of the tunnel barrier and in the local properties of the junction electrodes, which complicates the interpretation. Therefore, in this work we focused mainly on the variations of the energy gap, $\Delta$, of tunnel junctions in integrated circuits because they directly reflect changes in local properties of Nb electrodes.

## 3. Fabrication of NbAl/$AlO_x$/Nb Josephson junctions and diagnostic structures

The fabrication process with 4.5 kA/$cm^2$ critical current density was used for this work. The sequence of the layers and the fabrication details are given in [1], [2]. Series arrays of same-size circular JJs were used as test vehicles. To study the M1-GND junctions, the base electrode of the last junction in the array was connected to the ground plane by M1 layer via I0 contact hole. In the second group of arrays, the counter electrode of the last junction in the array was connected to the ground plane using M2 wiring layer (M2-GND junction). The length of this M2 wiring was varied. In the third group, the counter electrode of one or several junctions was connected to chip contact pads on the chip border. Each pad present an isolated metal structure containing a stack of all Nb layers in the circuit M0, M1, M2 and M3 covered by Ti/Au (30nm/200nm) or Ti/Pd/Au (30nm/100nm/200nm) contact pad metallization layer R3. The chip ground plane also has the contact metallization at the chip border. In all three groups, all other Nb leads to the arrays have molybdenum resistors inserted between the arrays and chip contact pads. The multi-JJ array arrangement allows us to see the deviation of the specially connected junction(s) from the rest of the junctions in the array, and at the same time establish the typical spread of the critical currents of the nominally identical junctions which do not have these special connections.

Figure 1 shows one of the test structures containing four floating junctions located in a circular fashion at equal distances to a contact hole to the ground plane [9]. The contact hole is 1.6 μm in diameter. The junctions are 1.9 μm in diameter. The distance between the centre of the contact hole and the centre of junctions was varied from 8 μm to 32 μm. Figure 2 shows the identical structure except that the contact hole is surrounded by a moat in the ground plane [9]. The total area of the ground plane layer around the contact hole in the first case is much larger than the area of base electrodes of the four junctions (and is about the total area of 5 x 5 $mm^2$ chip), whereas in the second case it is smaller than the total area of the junctions' base electrode.

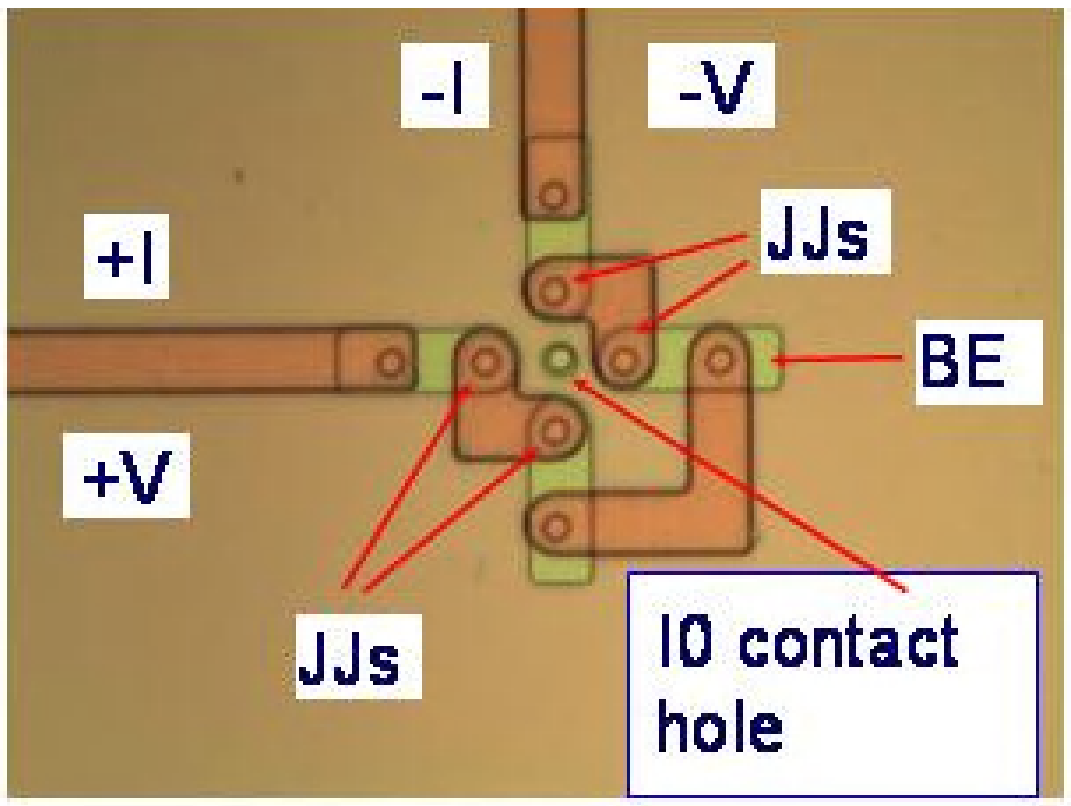

**Figure 1.** Series array of four junctions around contact hole, connecting junctions' base electrode (BE) to ground plane layer M0 during the time interval between trilayer deposition and BE etching. Leads and JJ wiring are in M2 layer.

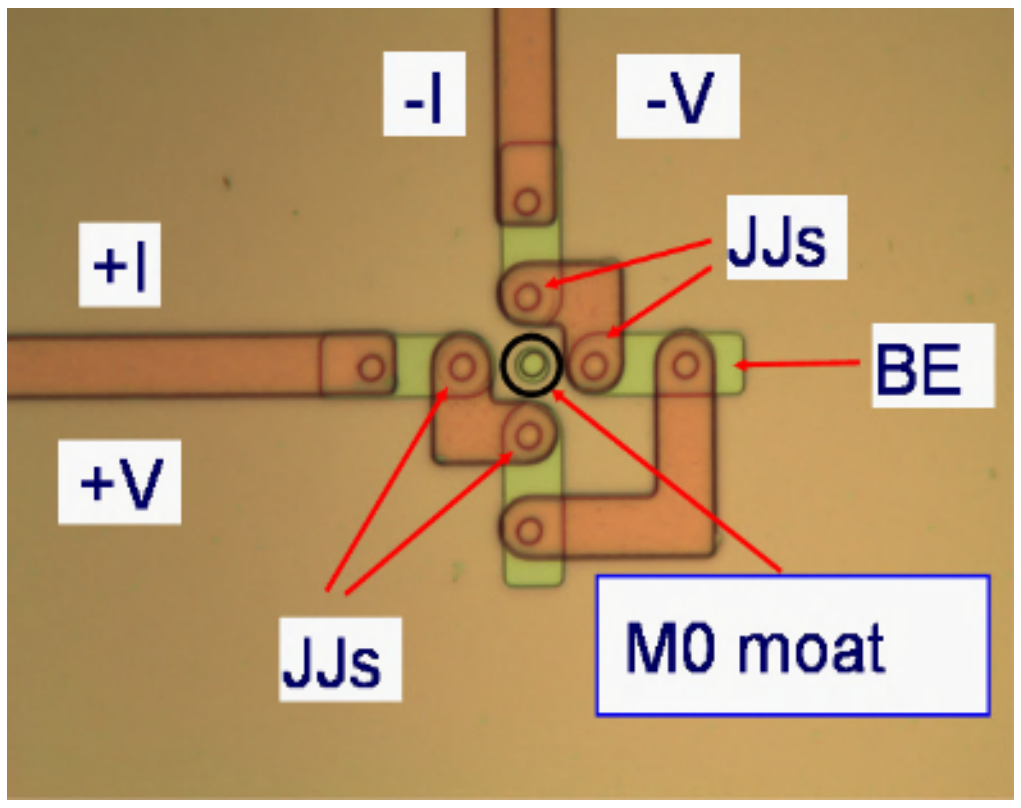

**Figure 2.** The same array of four junctions as in figure 1, but the contact hole is surrounded by a moat in M0. The moat separates the piece of ground plane inside the moat from the rest of the chip ground plane.

## 4. Experimental results

### *4.1. Effect of contact between junction BE and ground plane*

Figure 3 presents current-voltage characteristics of 4-junction arrays, shown in figure 1 and figure 2, at different distances between the centres of the junctions and the contact hole. It can be seen that as junctions become closer to the contact hole their critical current increases, indicating an increase in the average barrier transparency. At the same time, quality of their tunnel barrier decreases significantly as indicated by increase in subgap conductance and appearance of current steps at subharmonics of the gap voltage and excess current above the gap voltage $V$>2Δ. This effect was first observed in [9]. However, Josephson junctions in arrays shown in figure 2 demonstrate no influence of the contact hole either on their $I_c$ or on the barrier quality at any distances. Moreover, they demonstrate the highest barrier quality with respect to other junctions as indicated by more uniform critical current, lower subgap conductance, and higher gap voltage $V= n\cdot 2\Delta$ , $n$=4.

We remind that base electrodes of the junctions remain connected to the ground plane via the contact hole only from the time of Nb/Al/$AlO_x$/Nb trilayer deposition till the time of M1 (BE) layer patterning, after which base electrodes of different junctions separate from each other and from the ground plane. Therefore, etching fixes some state of Nb and tunnel barrier in each of now separate pieces of the base electrode, in which they evolved during the time from the deposition.

The dependence of the average critical current and of the average energy gap in Nb electrodes on the distance to contact hole is shown in figure 4 and figure 5, respectively. The junction gap voltage is the sum of Nb energy gaps in the base and counter electrodes. The counter electrodes in all arrays presented in figure 5 are identical and only the distance between the base electrodes and the contact hole changes. Therefore, changes in the gap voltage in figure 5 represent changes in the local gap of Nb base electrode caused by proximity to the contact to M0 layer. The most interesting is that the local Nb gap is the highest near the ground contact and changes with the distance, approaching the value of the gap in M0-moat-protected arrays (figure 2) or in arrays far away from the contact at distances ~ 40 to 50 μm.

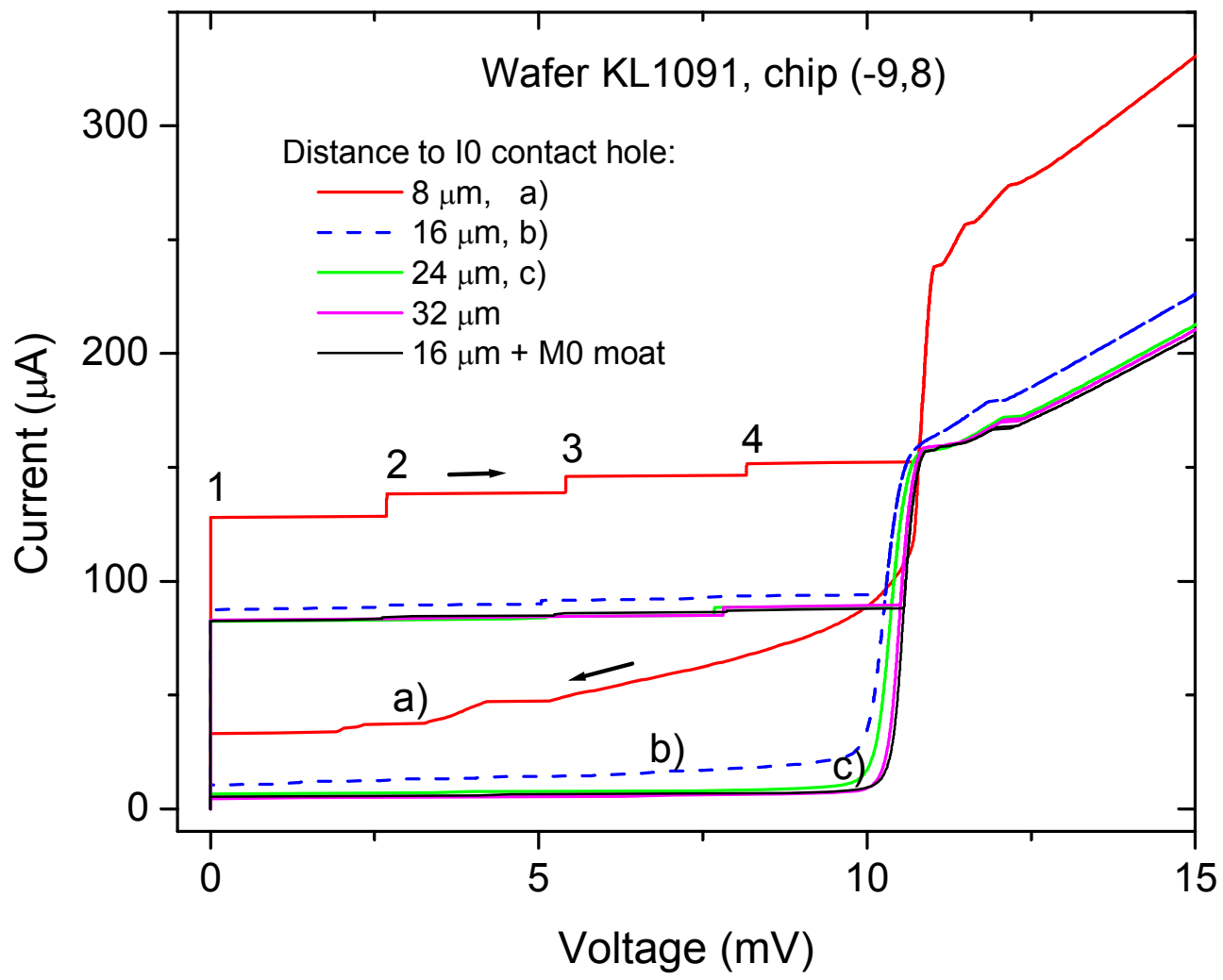

**Figure 3.** *I-V* characteristics of 4-JJ arrays around a contact hole to ground plane (see, figure 1 and figure 2). Arrays having a moat in ground plane around the contact hole (figure 2) demonstrate the highest quality of the tunnel barrier, the most uniform critical current, and the highest gap voltage.

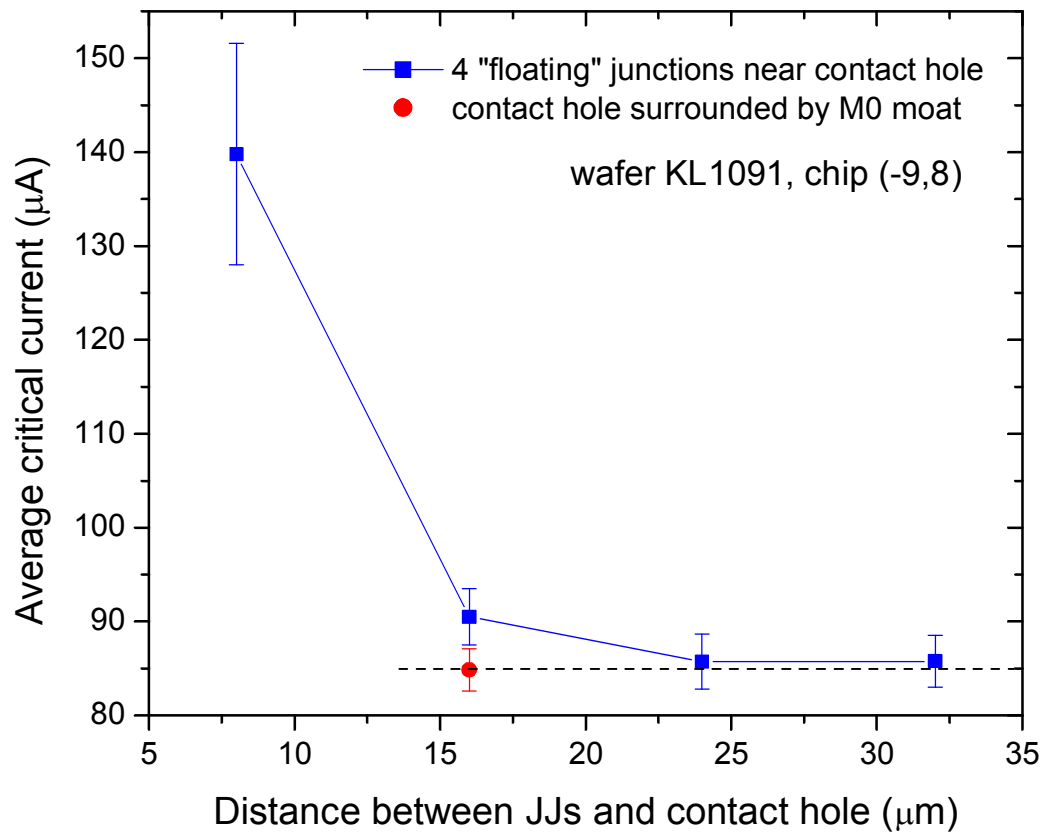

**Figure 4.** Average critical current of 4 "floating" JJs around a contact hole to ground plane, M0. Error bars show the spread of the critical currents. Data point connections in figure 4 and figure 5 are only to guide the eye.

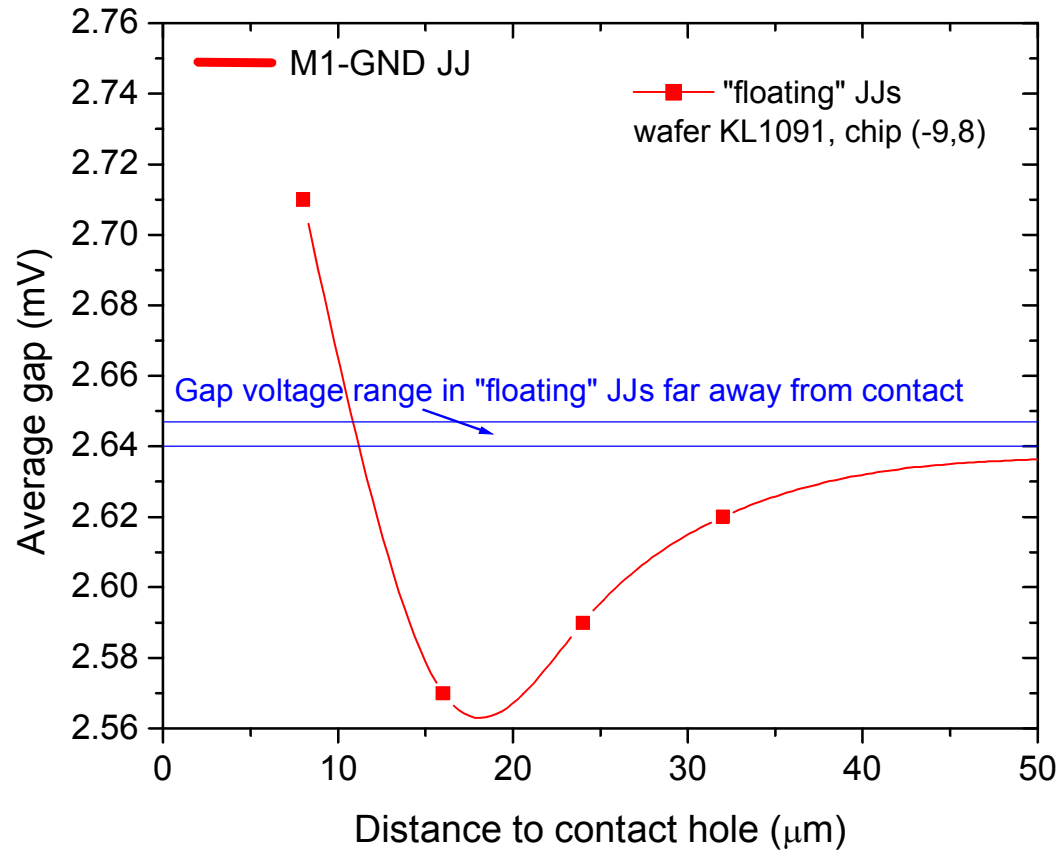

**Figure 5.** Average energy gap in 4 "floating"JJs around contact hole to ground plane M0. The range of gap voltages in arrays with M0 moat is shown as well as the gap in JJs with grounded base electrode (M1-GND).

For junctions with base electrode directly connected to the ground plane via I0 contact hole (M1-GND junctions), the effect on $I_c$ and the gap is very similar to the described above. The typical results are shown in figure 6 presenting *I-V* characteristics of a 20-junction array with the last junction grounded. It can be clearly seen that two junctions in the array have significantly higher critical current than the average and have much higher subgap conductance (damage to the tunnel barrier). These junctions have been identified as the last grounded JJ and its nearest neighbour, which are the two nearest junctions to the ground plane contact. This identification can be easily done by using extra leads connected to the junctions and allowing individual measurements of their *I-V* curves. For a comparison, we show the results for the same array surrounded by multiple contact holes to the ground plane. It was demonstrated first in [9] that surrounding JJs by multiple "dummy" contact holes to the ground plane layer has a protective effect and completely eliminates deviations of $I_c$ of the grounded junctions and damage to their tunnel barriers. As in the case of figure 1, these "dummy" holes become separated from the junctions after base electrode etching step of the processing.

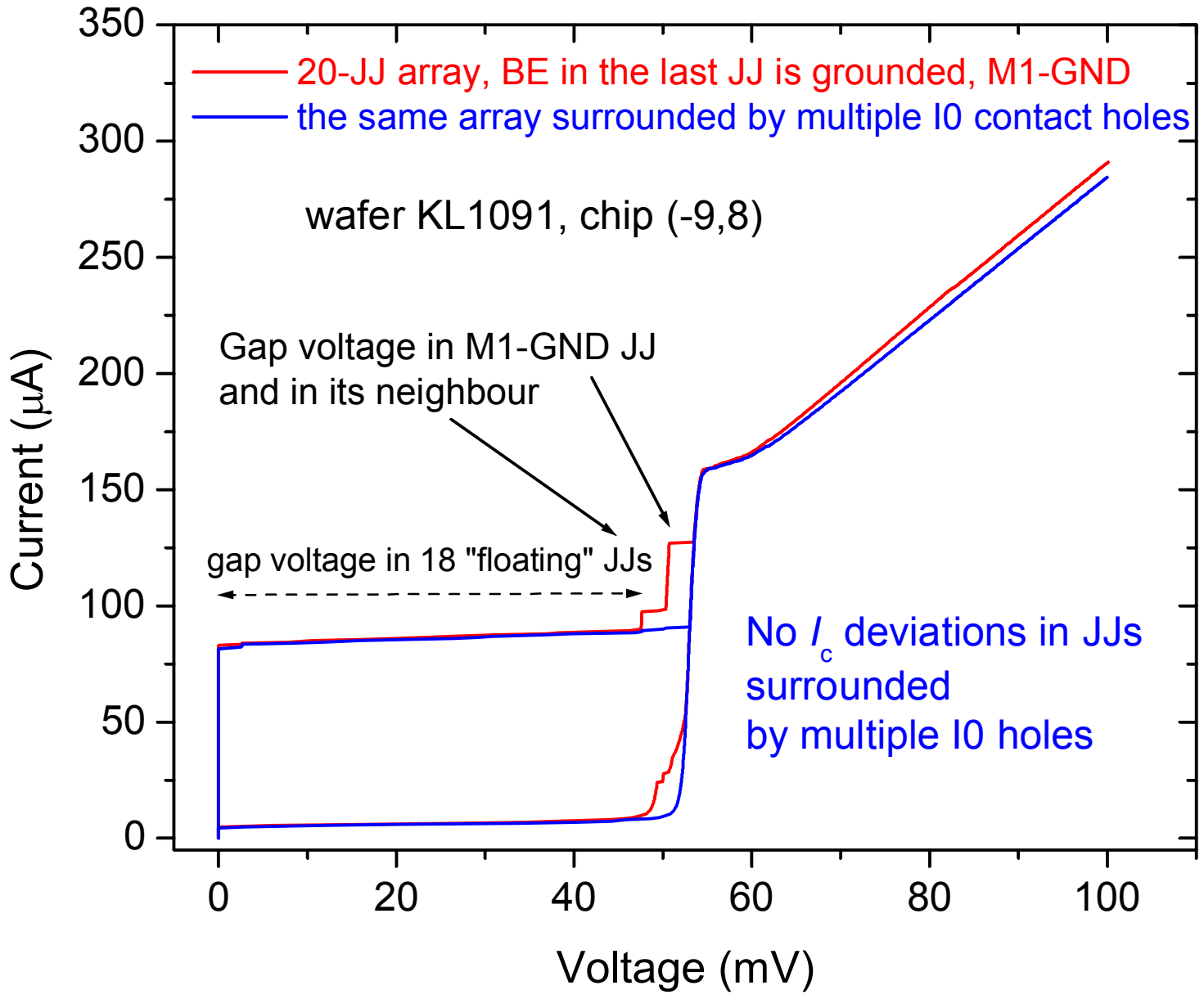

**Figure 6.** Current-voltage characteristics of 20-JJ arrays where the last junction in the array is connected to the ground plane by the base electrode layer. Two junctions showing the largest deviation in $I_c$ from the rest of junctions are the grounded junction and its near neighbour located in the proximity to the contact hole. The average gap in the first 18 "floating junctions is 2.64 mV, whereas the gap in the grounded junction and in its neighbour is 2.80 mV and 2.71 mV, respectively. The second array, showing no $I_c$ anomalies, is surrounded by multiple contact holes to the ground plane, working as a protective structure.

The average gap in the 18 junctions that are far away from the grounding contact can be easily found from the *I-V* curve, and is 2.64 mV. This is the same value as shown in figure 5 for the similar junctions. However, the gap in the grounded (20$^{th}$) junction is noticeably enhanced with respect to the "floating" junctions. Its value $\Delta_{M1\text{-}GND} = 2.80$ mV is close to the value of Nb gap observed in clean Nb films. The gap in the neighbouring 19$^{th}$ junction has an intermediate value of 2.71 mV since it is farther away from the ground contact, in agreement with the data in figure 5.

### *4.2. Effect of contact between junction CE and ground plane*

Another type of grounding is achieved by connecting junction counter electrode to the ground plane by M2 wire via I0 contact hole. To separate the effect of CE to GND connection from the shown in figure 3 effect of I0 contact hole, the I0 contact hole was placed far away from the junction (>>50 μm), so M2 wire was long. The typical *I-V* characteristics are shown in figure 7 for the twice larger area junctions than in the previous figures. In the case of M2-GND connection, no significant effect was found right after the wafer fabrication. However, when the same chip was re-measured ~1.5 years later, the grounded junction has shown a significant increase in both $I_c$ and the gap. Contrary to M1-GND junctions, no increase in subgap conductance or other signatures of tunnel barrier damage were found in case of M2-GND junctions. Measurements on other arrays not presented here have shown that $I_c$ enhancement in M2-GND JJs develops gradually on a time scale of several months and has a tendency to saturation in time. This clearly indicates a diffusion-related nature of this phenomenon.

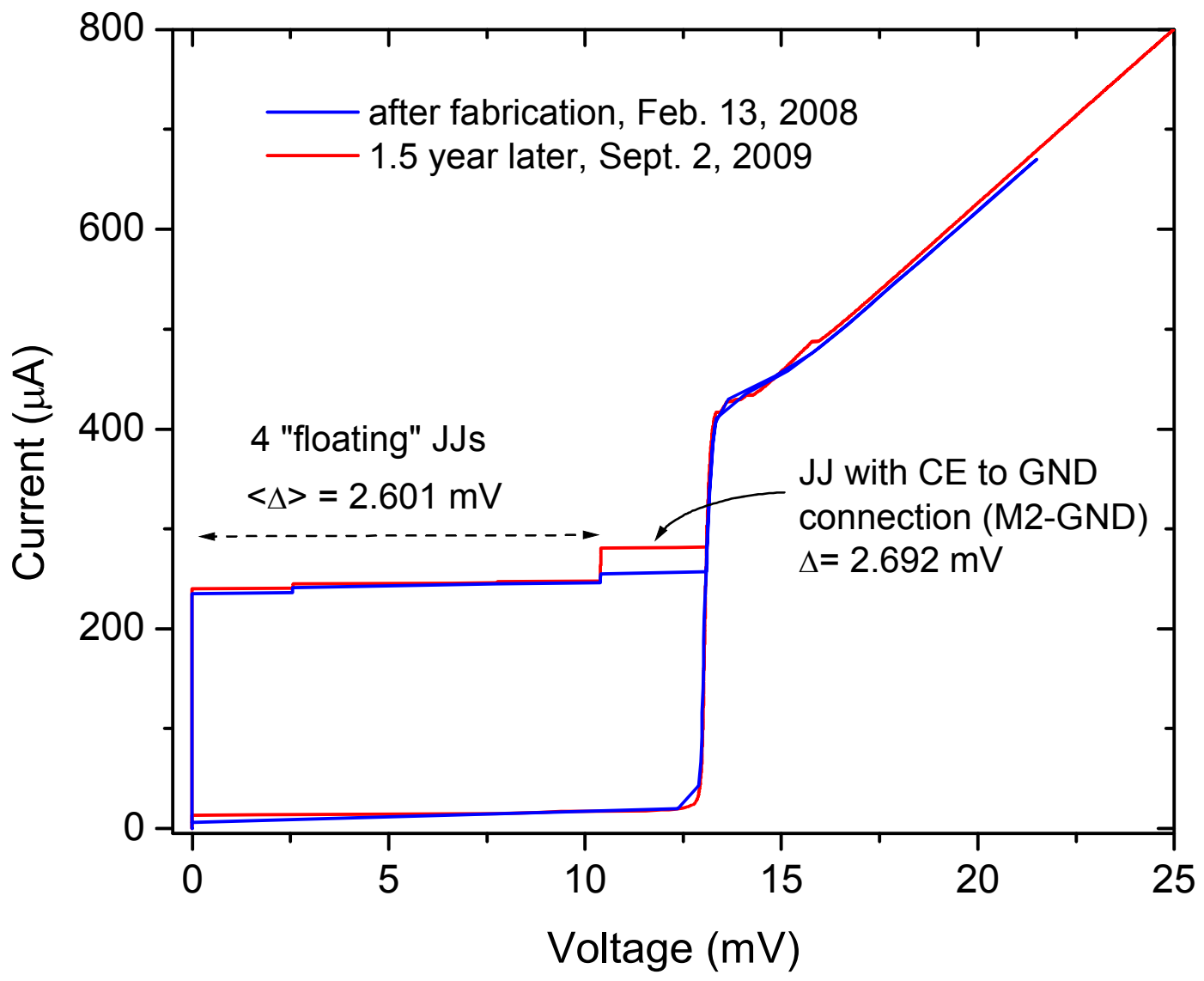

**Figure 7.** Five-junction series array of circular junctions with 2.66-μm diameter (design $I_c$=250 μA). In the last junction the counter electrode is connected to the ground plane layer M0 by a long M2 wire via I0 contact located far away from the junction. Right after the fabrication, the grounded junction shows no difference from the rest of JJs in the array. However, a significant $I_c$ enhancement has developed after long-term storage of the chip at room temperature without any degradation of the tunnel barrier quality.

*4.3. Effect of contact between junction CE and Ti/Au or Ti/Pd/Au contact pads*

Finally, we consider the effect of connection of junction counter electrode to chip contact pads (R3 layer) at the periphery of the chip by M2 wire (Nb). These pads present either Ti/Au bilayer or Ti/Pd/Au trilayer, with Ti serving as an adhesion layer. No difference in the effects below was noted whether Pd layer was used in the contact pad metallization or not. Figure 8 and figure 9 show changes in the current-voltage characteristics of a single JJ with CE to contact pad connection by M2 wire.

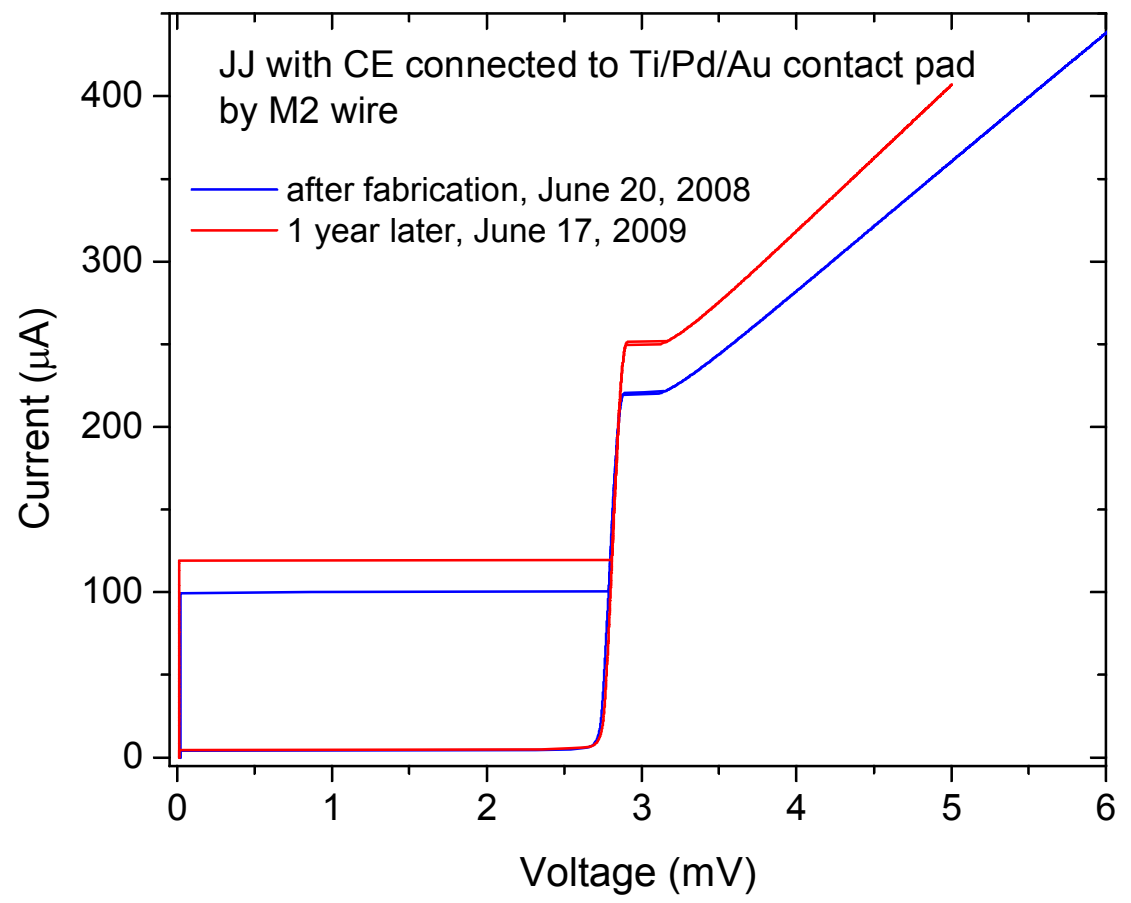

**Figure 8.** *I-V* characteristics of a single JJ with counter electrode connected to Ti/Pd/Au contact pad at the chip edge. The $I_c$ and the normal state conductance increase after long-term storage at room temperature. Wafer KL1095, da_chip03.

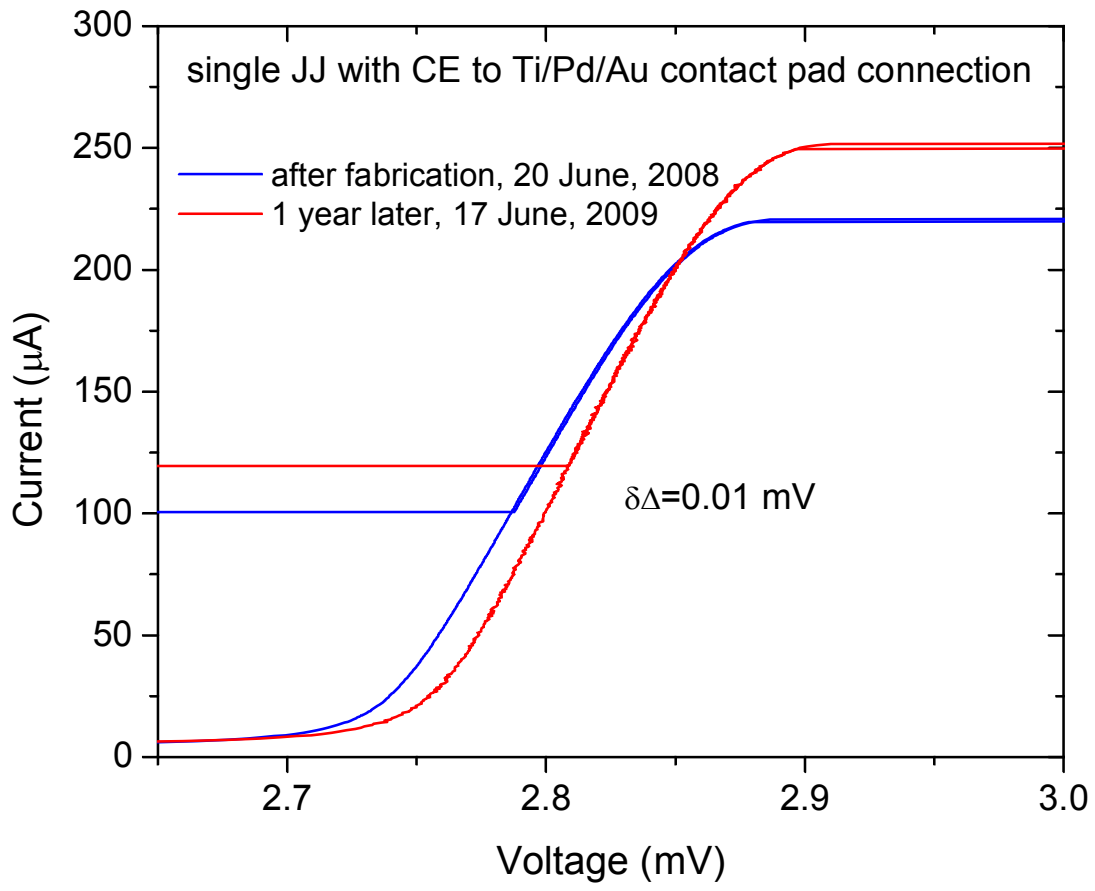

**Figure 9.** Blow-up of the gap voltage area of the same junction as in figure 8, showing a slight increase also in the gap voltage after storing the chip at room temperature for 1 year. Wafer KL1095, da_chip03.

It can be seen that there is a long-term effect of the contact pad connection on the junction which is similar to the above described effect of CE to GND connection by M2 wire. Namely, both the critical current and the normal state conductance increase after long-term storage (annealing) at room temperature. All the changes indicate that the overall junction quality increases because the superconducting gap in electrodes also slightly increases and the subgap conductance decreases. These changes look like Nb counter electrode of the junction becomes cleaner and closer to the ideal Nb. We note that the slight increase in the gap after storage cannot explain a much large change in the critical current, whereas the increase in $I_c$ and in the normal state tunnel conductance indicates that the tunnel barrier height decreases.

## 5. Discussion of the results

### *5.1. Hydrogen interlayer migration effect on Josephson junctions: the model*

The experimental results presented in the previous section can be explained in the following model. It is known that hydrogen can easily dissolve in Nb films during many processing steps [14]. Since different layers of Nb-based integrated circuit require different type of processing and different number of processing steps, they may end up with different concentration of dissolved hydrogen. For instance, the ground plane (M0) layer requires only Nb deposition and etching, whereas M1 layer processing involves a trilayer deposition, counter electrode etching, anodization, ion milling, and base electrode etching. So it is likely that, due to much heavier processing, M1 layer will have a different (likely higher) concentration of hydrogen than M0 layer. Due to a very high mobility of hydrogen in Nb [15], migration of hydrogen between M1 and M0 layers should take place through contact holes connecting these layers. Since M1 layer is a part of the base electrode of Josephson junctions, any changes in the concentration of hydrogen in M1 layer around these contact holes should affect Josephson junctions located within the diffusion length from the contact hole by a combination of the following mechanisms

- Increasing hydrogen concentration in Nb base electrode decreases somewhat the superconducting critical temperature [16], [17] and, hence, one can expect a decrease in the energy gap, $\Delta$ [16] and in the Josephson critical current.
- However, increasing of hydrogen concentration in Nb increases the lattice constant and increases the surface roughness [18]. This should stretch $AlO_x$ tunnel on Nb surface and promote formation of defects in the barrier, which should increase the barrier transparency, the critical current, and the subgap leakage current.
- Conversely, decreasing of hydrogen concentration in Nb base electrode should increase the gap, increase the critical current, decrease the lattice constant, and compress the barrier. This stress can also lead to the formation of defects and increase in the barrier transparency and $I_c$, and subgap conductance.

In contrast to M1-GND junctions, the floating junctions are not connected directly to the ground plane, and thus the change in hydrogen concentration in their base electrode does not occur (or is much smaller) that results in the lower critical current than in the M1-GND junctions. Therefore, in integrated circuits where both types of junctions (M1-GND and floating) are intermixed according to the logic design requirements, there can be large systematic deviations of one group of junctions (M1-GND) from the intended design values. Such deviations unavoidably reduce yield, margin of operation, and the maximum clock frequency. In floating junctions located in the proximity of I0 contact hole, $I_c$ can also become affected because hydrogen redistribution could have occurred during the time between the trilayer deposition and base electrode patterning, after which the junctions become floating and hydrogen redistribution between M1 and M0 layers is no longer possible.

The situation with M2 wiring is slightly different. Because M2 layer processing also involve only one deposition and one etching (similarly to M0), it is likely that it has different hydrogen concentration than the counter electrode (CE). However, connecting the junctions to M2 wiring should not create large differences between them because counter electrodes of **all** the junctions in a circuit become connected to M2 layer at the same time and remain connected from that point on. The difference between different JJs may only be in the area of the connected M2 wiring. However, the situation changes if M2 wiring to some junction connects also directly to M0 layer (we call this M2-GND junction), because a redistribution of hydrogen between M2 and M0 may then occur, causing some changes in properties of this junction.

When the next wiring layer M3 is deposited, it becomes connected only to some of the junctions. This again can create potential nonuniformity of the critical current if the properties of M3 layer (e.g., the hydrogen content) differ from other layers.

The presence of Ti/Au or Ti/Pd/Au layer in contact with Nb leads to the junction causes diffusion of hydrogen along Nb wires (and other impurities dissolved such as oxygen, nitrogen, carbon) towards Ti layer because it is a stronger getter of hydrogen than Nb. Therefore, Nb electrodes in proximity to the junction become cleaner, and electron mean free path as well as the density of states increase. As a result, the energy gap increases towards its value in clean Nb accompanied by an increase in $I_c$. By far the largest effect of hydrogen removal from the junction electrodes is on the normal state tunnelling conductance, because its increase (as well as the related increase in $I_c$) is the most significant change observed (see figure 8 and figure 7). This can be explained by a decrease in the average tunnel barrier height with reduction of hydrogen content in Nb. Indeed, the barrier height is determined by the difference between the work function in Nb and the bottom of the conduction band in $AlO_x$ barrier. According to photoelectron spectroscopy data, the work function of Nb increases by ~ 0.1 eV at hydrogen saturation [19]. Therefore, the average tunnel barrier should decrease with a decrease in the hydrogen content. The effect should be more pronounced on side of the counter electrode because it is adjacent to the oxide barrier, whereas the base electrode Nb is covered by 8-nm-thick Al layer. Therefore, the tunnel barrier height on the BE side is mostly determined by the work function of Al and does not change much because solubility of hydrogen in Al is very low and Nb is far away from the barrier.

Our experiments also indicate that migration of impurities between the base electrode and the ground plane leads to some structural damage to the tunnel barrier that reveals itself in increased subgap leakage current. The reason for this damage has not been established yet. In contrast, migration of impurities between the counter electrode and other layers (including the ground plane), although changes the critical current and normal state conductance, apparently does not produce structural damage to the barrier because the subgap conductance does not increase. This difference in the behaviour could be a result of the fact that M2 layer makes contact with the junction when it is completely sealed into $SiO_2$ interlayer dielectric providing significant mechanical stability to the junction, whereas open surfaces of Nb electrodes are free to deform between the moment of contact between BE and the ground plane and the moment of interlayer dielectric deposition over the counter electrode.

### *5.2. Detection of hydrogen in Nb films*

Hydrogen can easily dissolve in Nb when its surface is not protected by the oxide barrier. Therefore, hydrogen poisoning of Nb films can easily happen at many processing steps such as etching and ion milling when the surface oxide is removed. In order to monitor hydrogen content in Nb layers of integrated circuits and its changes during all processing steps, we implemented a qualitative method based on optical emission during reactive ion etching of Nb films in $SF_6$ plasma. During the etching, all chemical elements contained in the film are released into the vacuum chamber of the etcher and emit light due to various excitation and recombination processes in rf plasma. The intensity of lines of optical emission of hydrogen should be proportional to the amount of hydrogen dissolved in the initial Nb film, assuming that there are no other sources of hydrogen in the reaction chamber.

Emitted light was coupled via a fibre optic cable to a grating spectrograph, model SD1024 from Varity Instruments, Inc., [20]. We monitored the intensity of fluorine emission line at 703.7 nm for Nb etching end-point detection and the sum of intensities of hydrogen lines at 486.1 nm and 656.5 nm.

The typical time dependencies are shown in figure 10. Hydrogen emission lines are much weaker than fluorine line at 703.7 nm, so their intensity was scaled appropriately to be presented in the same graph.

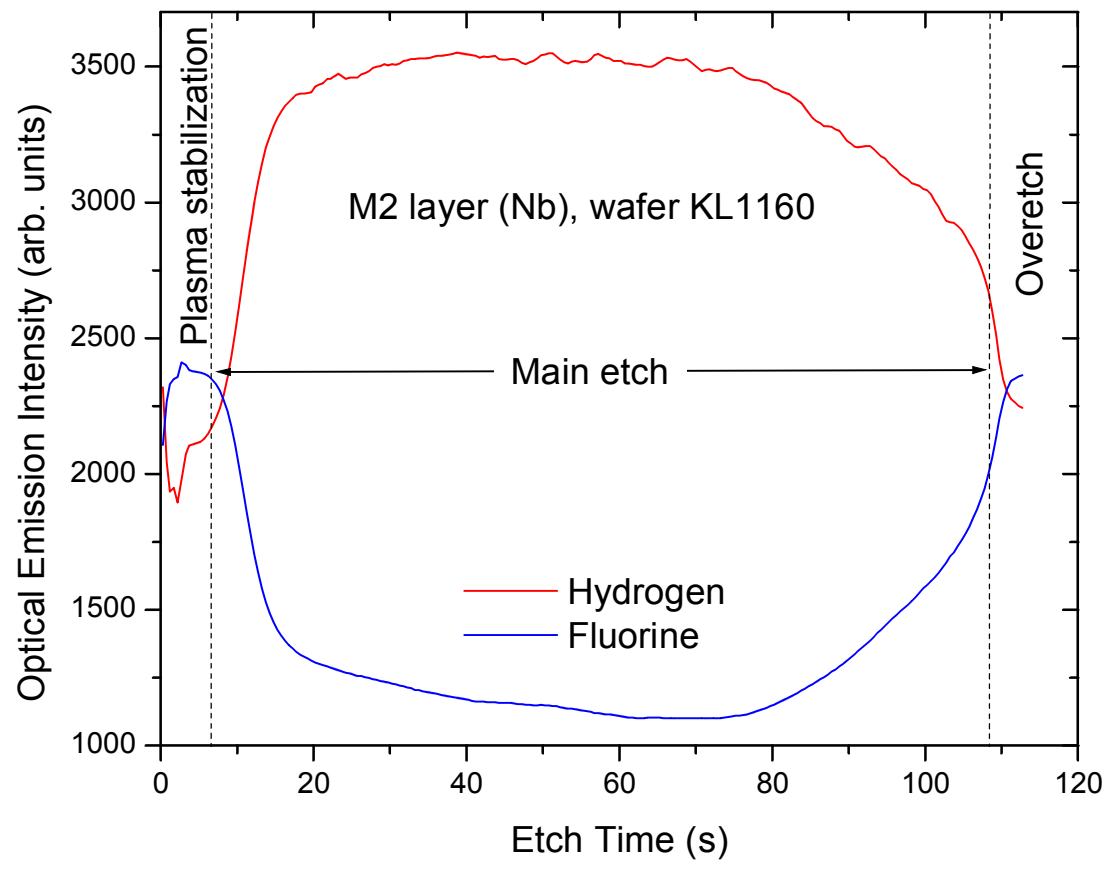

**Figure 10.** Intensity of F and H emission lines during Nb etching in $SF_6$ plasma. Nb consumes fluorine and releases hydrogen.

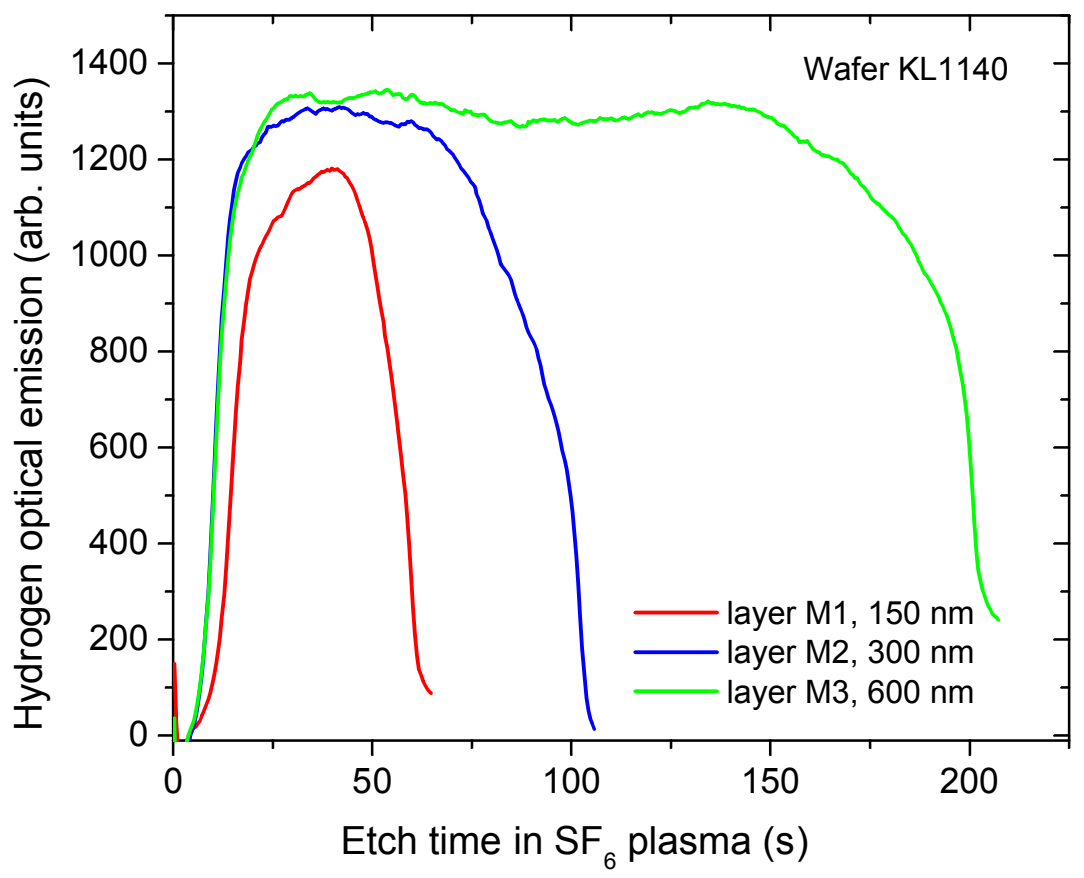

**Figure 11.** Intensity of H emission lines during etching of different Nb layers in $SF_6$ plasma. Background intensity was subtracted.

After a short period of plasma stabilization following the plasma ignition at *t*=0, the etching starts and the intensity of F-line starts to decrease due to fluorine consumption in the following reaction:
Nb (solid) + F (gas) → $NbF_5$ (gas). As can be seen in figure 10, at the same time the intensity of H-lines starts to increase from the constant background value, indicating an increase in the concentration of hydrogen in the plasma. Overall, there is a complete anti-correlation between F- and H- lines. After main etch of Nb is complete, the intensity of F-line restores to its initial value because there is no more Nb to react with, and the H-line intensity decreases to the background value because excess hydrogen released from Nb is pumped away from the reaction chamber while its supply ceases to exist. The amount of hydrogen released during the etching can be characterized by the integral of the H-lines intensity vs time curve after deducting the background intensity, which was taken as the intensity after plasma stabilization right before the start of the main etch.

Figure 11 shows the change in hydrogen optical emission during etching of the junction base electrode (M1), first wiring layer (M2), and second wiring layer (M3) on the typical wafer; the intensity normalization is the same as in figure 3, and the background intensity was subtracted. The etching was done at the identical plasma conditions. The thicknesses of the layers are different and scale as 1:2:4. So, the etching time increases proportionally. There is also a noticeable increase in the hydrogen emission from the thicker layers. This will need further monitoring and investigating.

## 6. Conclusions

To summarize, superconductor integrated circuits present multilayered structures in which Josephson junctions are connected to different Nb layers. If the concentration of impurities (e.g., hydrogen) in different layers is different, the migration and/or electromigration of impurities can take place due to concentration gradients, difference in the residual stress in the films, and electric potential differences induced by plasma processing. The resulting changes in the concentration of impurities depend on the details of connections to the junctions and circuit patterns. This redistribution of impurities affects local superconducting properties of Nb, creates local stresses in the tunnel barrier, and varies the barrier height, and thus induces local changes in the Josephson critical current density and, ultimately, local deviations of the critical current of Josephson junctions from the design values. Diffusion paths of hydrogen along Nb wires and through contacts between different layers can be disrupted by molybdenum resistors, $AlO_x$ tunnel barriers in the junctions and/or Nb oxide layers at the interfaces, all of which have low hydrogen solubility and mobility. This can create very complex distributions and gradients of impurity atoms in circuits, which will dependent on the circuit design and fabrication details, and hence could be difficult to predict and analyze.

We have demonstrated the existence of three distinct effects of junction connections on its properties in Nb-base Josephson junctions:

**a)** the effect of base electrode connection to the ground plane;
**b)** the effect of counter electrode connection to the ground plane and/or other Nb layers;
**c)** the effect of connection of junction electrodes to contact pad metallization which has Ti adhesion layer – a strong getter of hydrogen and other impurities.

The most significant for digital circuits is effect **a)** because BE-GND connection is usually accompanied by irreversible damage to $AlO_x$ tunnel barrier. Effects **b)** and **c)** cause long-term drift of the critical current and tunnel resistance of $Nb/Al/AlO_x/Nb$ junctions but do not degrade the junction quality. In particular, **c)** is responsible for the reversal effects of hydrogen absorption on $I_c$ observed in [13].

Clearly, the best way to avoid all the phenomena and effects described in this paper is to prevent hydrogen poisoning of Nb layers during the fabrication process. This, however, might be difficult to achieve in a reproducible manner. In this case, protective structures similar to those described in [9], which apparently homogenize the distribution of impurities, help to improve the margin of operation of complex RSFQ circuits.

**Acknowledgments**
Authors would like to thank Richard Hunt, John Vivalda and Daniel Yohannes for their participation in the fabrication process. We are also grateful to Alex Kirichenko for helping us in designing some of the test structures and to Deborah Van Vechten for her interest and support in this work. This work was supported by the US Navy Office of Naval Research grants N000140810224 and N000140910079.